\documentstyle[12pt]{article}
\def\QCD{\Lambda_{\overline{MS}}}
\def\OMT{ < \!\!  1-T \!\! > }
\def\MH{ < \!\! m_H^2/s \!\! > }

\begin{document}

\begin{titlepage}
\begin{flushright}
DTP/98/08\\
\end{flushright}
\begin{center}
{\Large\bf Determination of $\QCD^{(5)}$ from the measured energy
dependence of $< \!\!  1-{\rm Thrust} \!\! >$}\\
\vspace{1cm}
{\large
J.~M.~Campbell, E.~W.~N.~ Glover and C.~J.~Maxwell}\\
\vspace{0.5cm}
{\it
Physics Department, University of Durham,  Durham DH1~3LE, England} \\
\vspace{0.5cm}
\end{center}
\begin{abstract}
We directly fit the experimentally measured energy dependence of the
average value of $1-{\rm Thrust}$, $\OMT$, over the $e^+ e^-$
centre-of-mass energy range $Q=14 - 172$~GeV to the QCD
expectation obtained by
integrating up the evolution equation for \newline
$d\!\!\OMT\!\!/d\log Q$ in
terms of $\OMT$.  We fit for $\QCD^{(5)}$, uncalculated ${\cal
O}(\alpha_S^3)$ and higher perturbative corrections parameterized by
the scheme invariant $\rho_2$, and the parameter $\lambda$ which
characterizes the magnitude of the leading $1/Q$ power
corrections anticipated for $\OMT$. A 3-parameter fit yields
$\QCD^{(5)}=245^{+20}_{-17}$~MeV, $\rho_2=-16\mp 11$ and
$\lambda=0.27^{+0.12}_{-0.10}$~GeV, equivalent to
$\alpha_S(M_Z)=0.1194 \pm 0.0014$. In this approach, there is no error
associated with the choice of the renormalization scale $\mu$.
\end{abstract}
\end{titlepage}

For several $e^+ e^-$ QCD observables we now have experimental measurements
\cite{data,siggi} spanning the centre-of-mass energy range from the lowest
PETRA energy $Q=14~{\rm GeV}$ through LEP-I at $Q=91~{\rm GeV}$ up to LEP-II
at $Q=172~{\rm GeV}$.
An ideal observable for testing QCD is the average value of $1-{\rm Thrust}
\equiv  \OMT$, where the Thrust is defined to be,
\begin{equation}
T = \max \frac{\sum_k | \vec{p}_k\cdot\vec{n} |}{\sum_k |\vec{p}_k |},
\end{equation}
with the sum running over all particles in the event.
The thrust axis $\vec{n}$ is varied to maximize the thrust.
Thrust describes the jettiness of the event such that $T=1$ for events with
two back-to-back particles and $T=1/2$ for completely spherical events.
Since $T$ is fully inclusive, the averaging means that it is free of
the large kinematical logarithms which afflict distributions in jet
observables close to the two-jet region.

If we consider the observable $R(Q) \equiv \OMT \!\!/1.05$ then we
have a perturbation series and leading power correction of the form,
\begin{equation}
R(Q)=a+r_1 a^2+r_2 a^3 + \ldots +
\frac{\lambda}{Q} (1+\lambda_1a+\lambda_2a^2 + \ldots),
\label{eq:thrust}
\end{equation}
where $a \equiv \alpha_S(\mu)/\pi$ denotes the renormalization
group improved coupling.  Note that the normalization is simply such
that the perturbative expansion begins with unit coefficient. In the
${\overline{MS}}$ scheme with $\mu=Q$ and $N_f=5$ active quark
flavours the next-to-leading order (NLO) coefficient is $r_1 =
9.70$~\cite{nlothrust}. The next-to-next-to-leading order (NNLO)
coefficient $r_2$ is unknown.
Other uncalculated higher order corrections and genuine non-perturbative
corrections are included by the phenomenological
$1/Q$ power corrections~\cite{renormalon,renormalon2,Webber}.
In \cite{Webber},
the power corrections up to $\lambda_2$ are calculated in terms of a
scale $\mu_I$ representing the transition between the perturbative and
non-perturbative regimes.  To extract $\alpha_s(M_Z)$ from the data,
we just truncate the perturbative series for a given renormalization
scale $\mu = xQ$ (which is typically $x=1$), and a given value of
$\mu_I$ (typically 2~GeV).  Then, by comparing with experimental data,
we solve for $a$.  A recent analysis \cite{siggi} for $\OMT$ using
this approach finds,
\begin{equation}
\alpha_s(M_Z) = 0.1204 \pm 0.0013 \phantom{~}^{+0.0061}_{-0.0050} \phantom{~}
^{+0.0023}_{-0.0018},
\label{eq:sigresult}
\end{equation}
(with a $\chi^2/{\rm d.o.f}$ of $42.6/24$) where the first error is
purely experimental.  The second and third errors come from varying
the theoretical input parameters, first allowing $x$ between 0.5 and 2
and second for $\mu_I$ in the range $1-3$~GeV.  Clearly the estimate
of the theoretical error is dominated by the renormalization scale
uncertainty.

Alternatively, we may avoid the renormalization scale entirely and,
by differentiating Eq.~\ref{eq:thrust} with respect to $Q$ and
using the renormalization group equation for the running of $a$,
directly write an expression for the running of $R(Q)$ with $Q$ in
terms of $R(Q)$ itself~\cite{grunberg,ECother,chris1,chris2},
\begin{eqnarray}
\frac{dR}{d\log Q} &=& -b R^2(1+cR+\rho_2 R^2 + \ldots )
                       +K_0 R^{-c/b} e^{-1/bR}
                        (1+K_1 R + \ldots) +\ldots \nonumber \\
                   &\equiv& -b \rho(R).
\label{eq:running}
\end{eqnarray}
Here $b$ and $c$ are the first two universal terms of the
QCD beta-function,
\begin{eqnarray}
b&=&\frac{33-2N_f}{6},\\
c&=&\frac{153-19N_f}{12b}.
\end{eqnarray}
The quantity,
\begin{equation}
\rho_2 \equiv r_2+c_2-r_1c-r_1^2,
\label{eq:r2}
\end{equation}
is a renormalization scheme and renormalization scale (RS) invariant
combination of the NLO and NNLO perturbation series and beta-function
coefficients with, in the $\overline{MS}$ scheme,
\begin{equation}c_2 = \frac{77139-15099N_f+325N_f^2}{1728b}.\end{equation}
Since the NNLO $r_2$ is unknown, so is $\rho_2$. As we will see later,
the coefficient
$K_0$ is directly related to the coefficient $\lambda$ of the $1/Q$ power
corrections in eq.~(\ref{eq:thrust}).

Since $R(Q)$ and $dR/d\log Q$ are both observables one could
in principle directly fit eq.~(\ref{eq:running}) to the data and thus
constrain the unknown coefficients $\rho_2$ and $K_0$. At asymptotic
energies all observables run universally according to,
\begin{equation}
\frac{dR}{d\log Q} = -bR^2(1+cR),
\label{eq:univrun}
\end{equation}
and one could see how close the data are to this evolution
equation. Given the error bars of the data and the separation in $Q$
of the different experiments it is preferable, however, to integrate
up eq.~(\ref{eq:running}) using asymptotic freedom ($R(Q) \to 0$ as $Q
\to \infty$) as a boundary condition.  In this way one
obtains~\cite{chris1},
\begin{equation}
\frac{1}{R }+c \log \left( \frac{cR}{1+cR} \right)
 = b \log\left(\frac{Q}{\Lambda_R}\right)
  - \int_0^{R} dx \, \left(-\frac{1}{\rho(x)}+\frac{1}{x^2(1+cx)} \right),
\label{eq:intup}
\end{equation}
where $\Lambda_R$ is a constant of integration. By comparing with the
$Q \to
\infty$ behaviour of eq.~(\ref{eq:thrust}) one can deduce
that~\cite{chris1},
\begin{equation}
\Lambda_R = e^{r/b} \left(\frac{2c}{b}\right)^{-c/b} \QCD,
\label{eq:RtoMS}
\end{equation}
where $r \equiv r_1^{\overline{MS}}(\mu = Q)$. Evaluating this for
$<1-T>$ yields $\Lambda_R = 14.4 \QCD$.

If we assume that the right-hand side of eq.~(\ref{eq:running}) is
adequately parameterized by,
\begin{equation}
-b\rho(R)=-bR^2(1+cR+\rho_2R^2)+K_0 R^{-c/b}e^{-1/bR},
\label{eq:param}
\end{equation}
we can then insert this form into eq.~(\ref{eq:intup}) and by
(numerically) solving the transcendental equation, perform fits of
$\rho_2$, $K_0$ and $\QCD$ to the data $R(Q)$. The fitted $K_0$ can
then be converted into an estimate of the parameter $\lambda$ in
eq.~(\ref{eq:thrust}) by differentiating eq.~(\ref{eq:thrust}) with
respect to $\log Q$ and comparing terms. We find,
\begin{equation}
\lambda = -K_0 \, e^{r/b} \left(\frac{b}{2}\right)^{c/b} \QCD .
\label{eq:lambda}
\end{equation}

\begin{figure}[t]\vspace{18cm}
\includegraphics{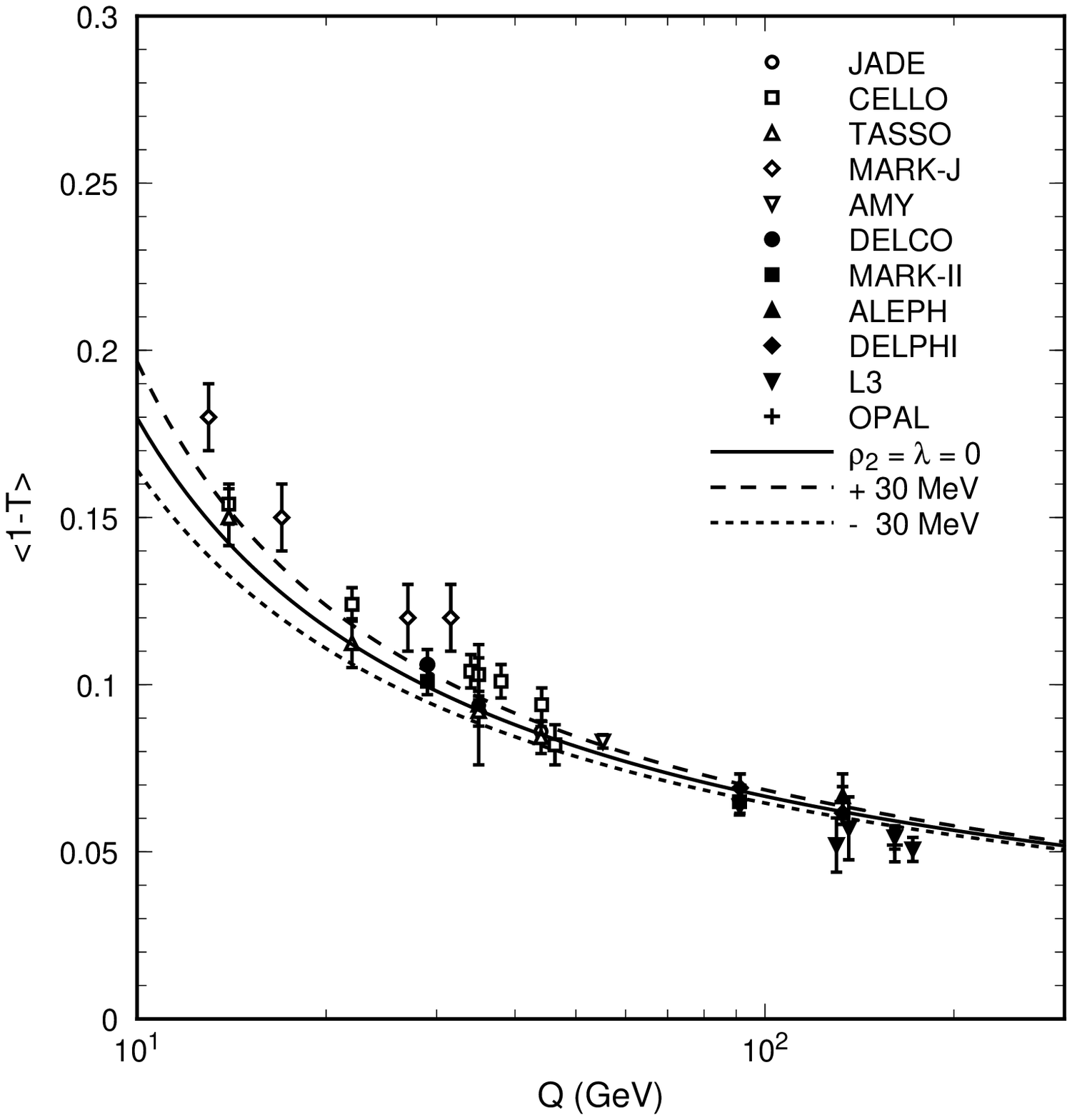}
\caption[]{The average value of $1-T$ obtained experimentally
\cite{data,siggi}
compared with the QCD expectation of eq.~(\ref{eq:intup}).
The solid line shows the fit to the data with $\rho_2 = \lambda = 0$
corresponding to $\QCD = 266$~MeV.
The long-dashed and short-dashed lines show the effect of altering $\QCD$ by
$\pm 30$~MeV.}
\label{fig:1parfit}
\end{figure}

\begin{figure}[t]\vspace{18cm}
\includegraphics{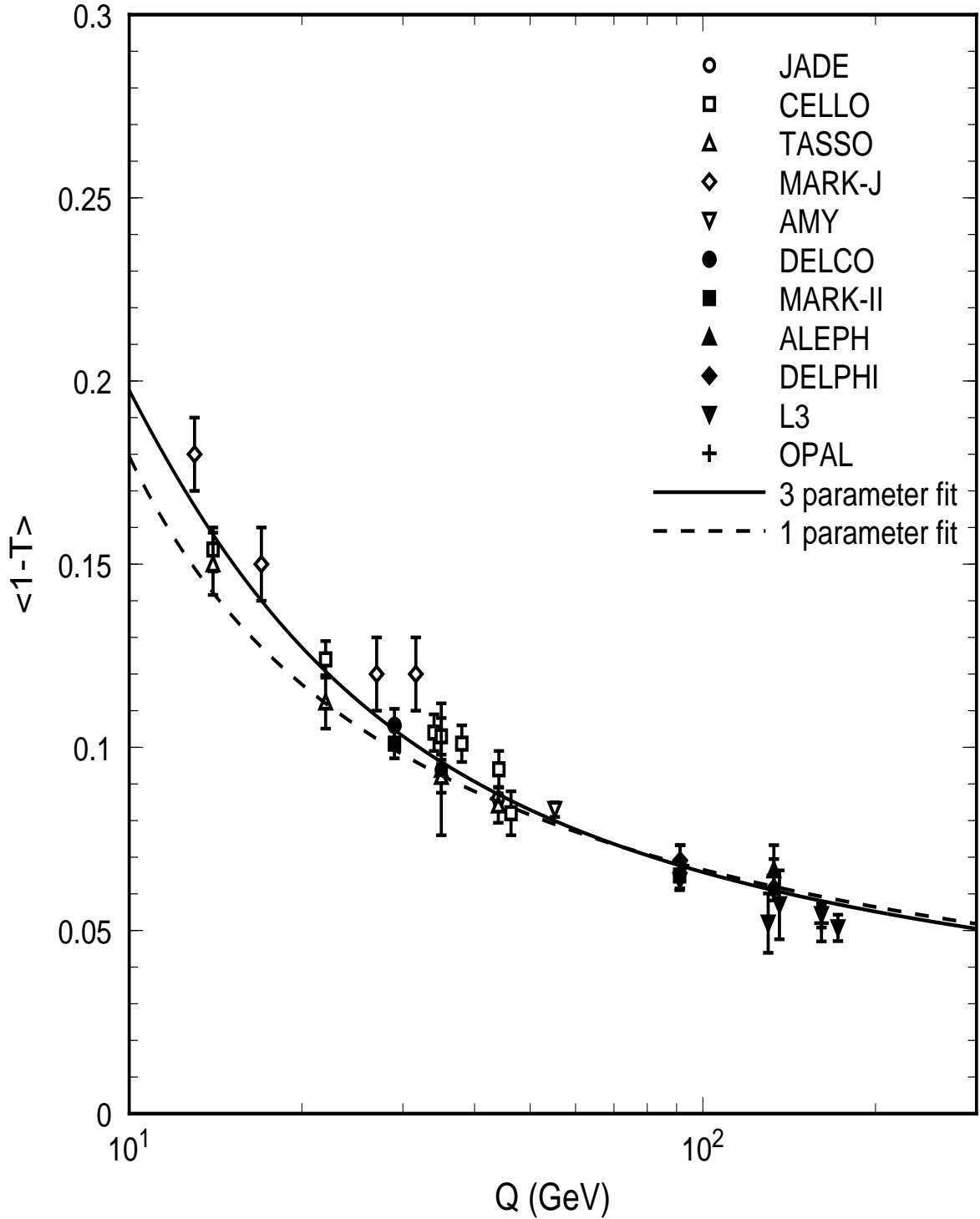}
\caption[]{The average value of $1-T$ obtained experimentally
\cite{data,siggi}
compared with the QCD expectation of eq.~(\ref{eq:intup}).
The dashed line shows the fit to the data with $\rho_2 = \lambda = 0$
while the result of the three parameter fit (to $\QCD, ~\rho_2$ and $\lambda$)
is shown as a solid line.}
\label{fig:3parfit}
\end{figure}

In Fig.~\ref{fig:1parfit} we show the fit to the data obtained by
setting $\rho_2=\lambda=0$. This corresponds to the universal running
of the observable given in eq.~(\ref{eq:univrun}). The fitted value is
$\QCD^{(5)} = 266~{\rm MeV}$ with a $\chi^2/{\rm d.o.f} = 81.7 / 32
$. The dashed lines show the effect of varying $\QCD$ by $\pm 30~{\rm
MeV}$ around the fitted value.  We clearly see that the data is
falling much too quickly with increasing $Q$ for the asymptotic
behaviour to have set in at these scales.  The data favours a more
steeply falling evolution which could be caused by either higher order
corrections with a positive $\rho_2$, or power corrections with
non-zero $K_0$.  We therefore perform a 3-parameter fit allowing
$\rho_2$, $K_0$ and $\QCD$ to vary independently which is shown in
Fig.~\ref{fig:3parfit}.  The minimum
$\chi^2$ fit is acceptable ($\chi^2/{\rm d.o.f} = 40.4 / 30 $) and
estimating an error by allowing $\chi^2$ within $1$ of the minimum
gives,
\begin{displaymath}
\QCD^{(5)}=245 ^{+20}_{-17}~{\rm MeV}
\qquad {\rm with} \qquad
\rho_2=-16\mp 11
\qquad {\rm and} \qquad
\lambda=0.27^{+0.12}_{-0.10}~{\rm GeV}.
\end{displaymath}
These values of $\rho_2$ and $\lambda$ are reasonably small, thereby
lending support to our critical assumption that the evolution equation
could be parameterized in this way.  We can convert the extracted
value of $\QCD$ into $\alpha_S(M_Z)$ using,
\begin{equation}
\alpha_s(M_Z) = \frac{2\pi}{b\log(Q^2/\QCD^2)}
\left [ 1-\frac{2c\log(\log(Q^2/\QCD^2))}{b\log(Q^2/\QCD^2)}\right ],
\end{equation}
and find,
\begin{equation}
\alpha_S(M_Z)=0.1194 \pm 0.0014.
\end{equation}
The fitted $\rho_2$ value may also be converted, if desired, into an
estimate of $r_2^{\overline{MS}} (\mu = Q)= 89\pm 11$.  Note
that this large value of $r_2$ is almost entirely due to the
renormalization group predictable $r_1^2+cr_1$ piece relating $\rho_2$
and $r_2$ (see eq.~(\ref{eq:r2})).

We see that our central value is remarkably close to that obtained by
\cite{siggi}.  The main difference is in how the errors are
determined.  In our approach, the errors are estimated by allowing the
uncalculated higher orders to be fitted by the data and the data
prefers these to be small.  As higher order corrections become known,
the new RS-invariant terms, $\rho_2$, $\rho_3$ etc., can be
incorporated and the fit refined.

\begin{figure}[t]\vspace{18cm}
\includegraphics{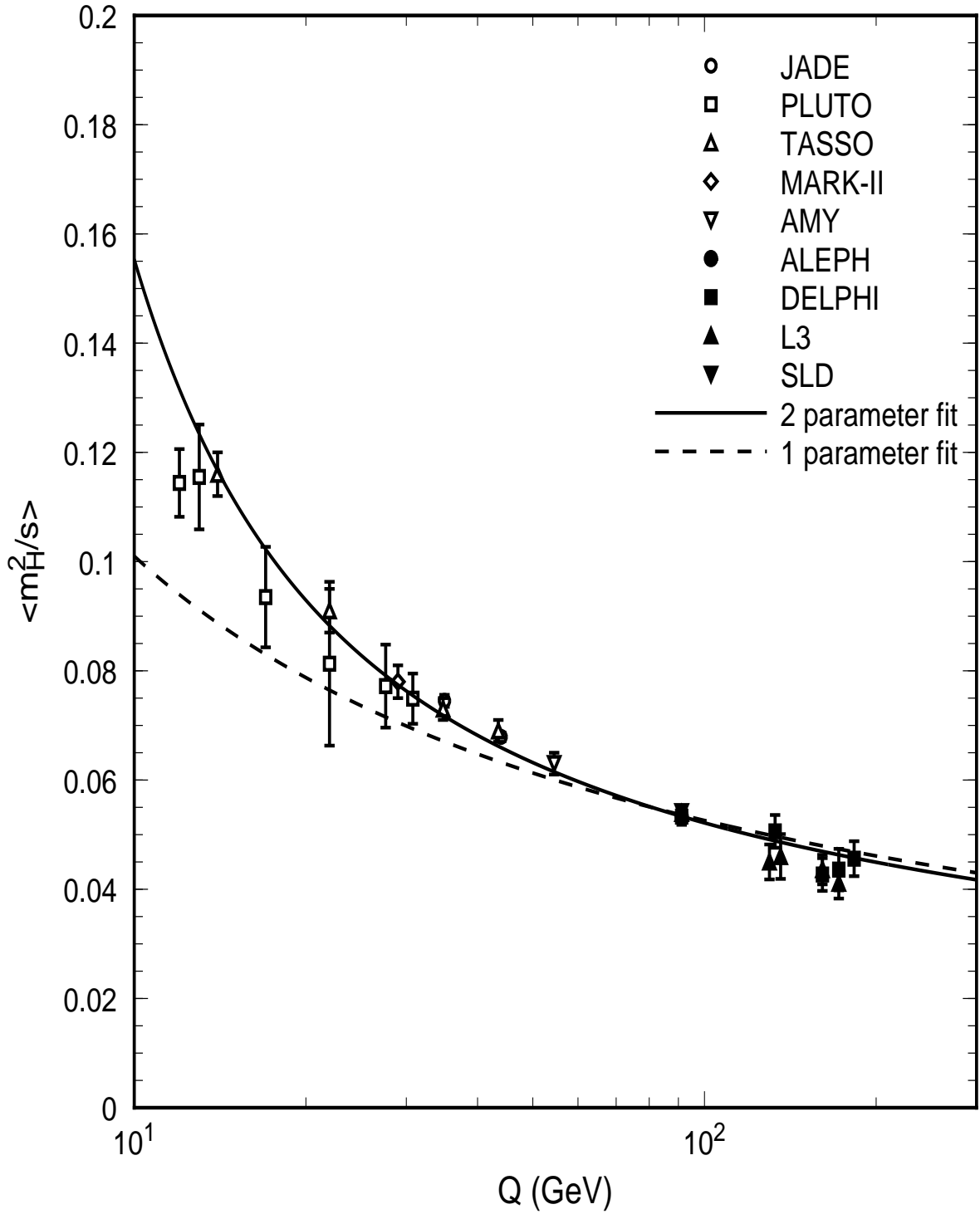}
\caption[]{The average value of the heavy jet mass
obtained experimentally
\cite{data,siggi}
compared with the QCD expectation of eq.~(\ref{eq:intup}).
The dashed line shows the fit to the data with $\rho_2 = \lambda = 0$
while the result of the
two parameter fit (to $\rho_2$ and $\lambda$) using the
value of $\QCD$ obtained from the three parameter fit to $\OMT$
is shown as a solid line.}
\label{fig:2parfit}
\end{figure}

With such an accurate value of $\alpha_S(M_Z)$, we should expect that
applying this approach to other variables should yield consistent results.
Unfortunately, the method described here relies on having
reliable data over a reasonable range of $Q$ values.
Another variable for which a wide range of data has been accumulated
is the heavy jet mass,
which is obtained by assigning the particles to one of two
hemispheres $H_1$ and $H_2$
according to whether $\vec{p}_k\cdot \vec{n}$ is positive or negative,
and finding the maximum scaled invariant mass of the hemispheres,
\begin{equation}
\frac{m_H^2}{s} = \frac{1}{s}  \max_{i=1,2} \left(\sum_{k\in H_i}
p_k\right)^2.
\end{equation}
Here $R(Q) = \MH \!\!/ 1.05$ while $r_1 = 4.52$~\cite{nlothrust}.
If we repeat the same analysis for this variable,  we find that
a one parameter fit with $\rho_2 = \lambda = 0$
gives a very poor fit, $\chi^2/{\rm d.o.f} = 213/29$ and $\QCD = 368$~MeV.
As seen in Fig.~\ref{fig:2parfit}
the data evolves much faster than the QCD prediction.
However, allowing both $\rho_2$ and $\lambda$ to vary while using the
value of $\QCD = 245$~MeV obtained from the $\OMT$ analysis gives a much
more satisfactory description.\footnote{Unfortunately,
in a three parameter fit,
$\rho_2$ and $\lambda$ trade off against each other and drive $\rho_2$ and
$\lambda$ to unacceptably large values
where we would have no reason to believe that
the parameterization is adequate.}
Here, $\chi^2/{\rm d.o.f} = 40.3/27$ while $\rho_2 = 13$ and
$\lambda = 0.11$~GeV  are sufficiently
small to support our choice of parameterization.

It is clear that the only certain way to reduce theoretical errors is
to compute the NNLO coefficients $\rho_2$.
However, fitting directly to
the energy dependence of the data as we have done enables us to
estimate them together with possible power corrections.  The
renormalization group scale-dependent logarithms are automatically
resummed to all orders on integrating eq.~(\ref{eq:running}) and do
not add a spurious extra large uncertainty in the extraction of $\QCD$
(or equivalently $\alpha_s(M_Z)$).
As a result, we have obtained a value of the strong coupling constant
that is both competititive with and in agreement with the
current world average \cite{pdg} of a wide range of data,
$$
\alpha_S(M_Z) = 0.119 \pm 0.002.
$$

{\noindent {\bf Acknowledgements}}

We thank Otmar Biebel for clarifying some of the values of
the experimental data points used in Ref.~\cite{siggi} and Daniel Wicke for
communicating the recent DELPHI measurements the heavy jet mass
of Ref.~\cite{data}.
We also thank Matthew Cullen for helpful comments
in the early stages of this work.
JMC thanks the UK Particle Physics and Astronomy Research Council
for the award of a research studentship.

\end{document}